\documentclass[10pt,conference]{IEEEtran}
\IEEEoverridecommandlockouts
\usepackage{authblk}
\usepackage{cite}
\usepackage{amsmath,amssymb,amsfonts,dsfont}
\usepackage{algorithm}
\usepackage[noend]{algorithmic}
\usepackage{graphicx}
\usepackage{textcomp}
\usepackage{xcolor}
\usepackage{bm}
\usepackage{booktabs}
\usepackage{multirow}
\usepackage{lipsum}
\usepackage{url}

\ifCLASSOPTIONcompsoc
\usepackage[caption=false, font=normalsize, labelfont=sf, textfont=sf]{subfig}
\else
\usepackage[caption=false, font=footnotesize]{subfig}

\newcommand{\thatis}{\textit{i.e.}}

\newcommand{\mbbR}{\mathbb{R}}

\newcommand{\bdw}{\boldsymbol{w}} 
\newcommand{\bdr}{\boldsymbol{r}}

\newcommand{\bda}{\boldsymbol{\alpha}} 
\newcommand{\bdb}{\boldsymbol{\beta}}

\newcommand{\mcX}{\mathcal{X}}
\newcommand{\mcP}{\mathcal{P}}
\newcommand{\mcY}{\mathcal{Y}}


\begin{document}

\title{A Communication Optimal Transport Approach to the Computation of Rate Distortion Functions 
\thanks{The first two authors contributed equally to this work.}
\thanks{$\dag$ Corresponding authors.}
}

\author[1,2]{Shitong Wu}
\author[1,2]{Wenhao Ye}
\author[1$\dag$]{Hao Wu}
\author[2$\dag$]{Huihui Wu}
\author[3$\dag$]{Wenyi Zhang}
\author[2]{Bo Bai}
\affil[1]{Department of Mathematical Sciences, Tsinghua University, Beijing 100084, China}
\affil[2]{Theory Lab, Central Research Institute, 2012 Labs, Huawei Tech. Co. Ltd., Hong Kong SAR, China}
\affil[3]{Department of Electronic Engineering and Information Science, 
\authorcr University of Science and Technology of China, Hefei, Anhui 230027, China 
\authorcr
Email: hwu@tsinghua.edu.cn, wu.huihui@huawei.com, wenyizha@ustc.edu.cn}

\maketitle

\begin{abstract}
In this paper, we propose a new framework named Communication Optimal Transport (CommOT) for computing the rate distortion (RD) function. 
This work is motivated by observing the fact that the transition law and the relative entropy in communication theory can be viewed as the transport plan and the regularized objective function in the optimal transport (OT) model. 
However, unlike in classical OT problems, the RD function only possesses one-side marginal distribution. 
Hence, to maintain the OT structure, we introduce slackness variables to fulfill the other-side marginal distribution and then propose a general framework (CommOT) for the RD function.
The CommOT model is solved via the alternating optimization technique and the well-known Sinkhorn algorithm.
In particular, the expected distortion threshold can be converted into finding the unique root of a one-dimensional monotonic function with only a few steps. 
Numerical experiments show that our proposed framework (CommOT) for solving the RD function with given distortion threshold is efficient and accurate.
\end{abstract}


\section{Introduction} \label{sec_introduction}
The rate distortion (RD) theory, first introduced by Shannon in \cite{shannon1948mathematical,shannon1959coding}, is a fundamental concept in information theory.
It is used to characterize the trade-off between the lossy compression rate and the distortion threshold.
Nowadays, it has a wide range of applications, such as image and video compression \cite{model_skodras2001jpeg,model_wiegand2003overview} and semantic information theory \cite{zhang2021rate}.
Despite the great importance of the RD theory, the analytic expression of the RD function is only known for some typical distributions with specific distortion measures \cite{berger1971,book_element}, making it difficult to solve real problems of interest.  
%
%
To date, the most well-known numerical method for computing the RD function is the Blahut-Arimoto (BA) algorithm \cite{blahut1972computation,arimoto1972algorithm}, which solves the classical RD function by alternating minimization.
Unlike the BA algorithm and its variants\cite{BA_csiszar1974computation}, we propose a novel framework based on the optimal transport (OT) approach for quantitative measurement of bounds in communication theory, which we name as the Communication Optimal Transport (CommOT). 
The OT problem is a famous mathematical problem with long history \cite{bogachev2012monge}. 
The proposal of the CommOT framework in the present paper is inspired by the connection between bounds in communication theory and the OT problem. 
%
%
Specifically, the transition law and relative entropy function in communication theory \cite{bai2020information} are regarded as the transport plan and objective function in the entropy regularized OT problem. 
Also, the distributions of the source and reproduction alphabets in the former are viewed as the marginal distribution constraints in the latter. 
Recently, GPU acceleration and the Sinkhorn algorithm \cite{cuturi2013sinkhorn}, \cite{2022wth} have greatly boosted the computation of the regularized OT problem, which drives us to consider the application of the OT approach in communication theory.
In our previous work \cite{ye2022optimal}, we proposed an efficient method based on the Sinkhorn algorithm for solving the Lower bound to the Mismatch capacity (LM rate) \cite{2000Mismatched} by observing its connection with the OT problem.
However, compared with the LM rate, computing the RD function is more challenging since it cannot be directly written as a similar formulation to the OT problem due to, most critically, lack of one-side marginal distribution and complexity introduced by the mutual information term in the objective function.
%
%
Thus, we introduce slackness variables to fulfill the other side marginal distribution, simplifying the objective function and developing a consistent formulation of the CommOT model.

Following this way, we are able to design an algorithm based on the Sinkhorn algorithm and the alternating optimization method to solve the CommOT model with high efficiency, named herein as the Alternating Sinkhorn (AS) algorithm.
Unlike the BA algorithm, the AS algorithm is designed for solving the RD function specifically under the CommOT framework. 
The advantages of the AS algorithm mainly lie in the following two aspects: 
(i) Given a distortion $D$, the corresponding rate $R$ can be obtained directly; 
(ii) When the rate distortion curve has a linear segment with a constant slope, the curve of the line segment can still be obtained directly. 
Numerical studies show that for the cases of binary, Gaussian, and Laplacian sources, our proposed CommOT model and the AS algorithm are robust and accurate.
The remaining part of this paper is organized as follows.
In Section \ref{sec_formulation_2},  we briefly review the basic definition of the rate distortion function and propose the CommOT model.
Next, we present the Alternating Sinkhorn algorithm in Section \ref{sec_sinkhorn_3} to solve the proposed CommOT model. 
In Section \ref{sec_num_4}, the simulation results demonstrate the advantages of our model and method.
We finally conclude the paper in Section \ref{sec_conclu_5}.
%


\section{Communication Optimal Transport Model} \label{sec_formulation_2}
We consider a memoryless source $X\in\mcX$ with reproduction $Y\in\mcY$, where $\mcX=\{x_1,\cdots,x_M\}, \mcY=\{y_1,\cdots,y_N\}$ are finite and discrete alphabets. 
Given a marginal distribution $P_{X}\in\mcP_{\mcX}$ and a distortion measure $d:\mcX\times\mcY\rightarrow\mbbR^{+}$, the classical RD function is defined as  \cite{berger1971,book_element},
\begin{equation}\label{rate_dis_tor_def}
R(D):=\min_{W(y \mid x): \sum_{x,y} P_{X}(x) W(y \mid x) d(x, y) \leq D} I(X ; Y).
\end{equation}
Here $I(X ; Y)$ is the mutual information between $X$ and $Y$, and $D$ is the distortion threshold.
Since the alphabets $\mcX$ and $\mcY$ are discrete and finite, we denote $w_{ij}=W(y_{j}\mid x_{i}), p_{i}=P_{X}(x_{i})$ and $d_{ij}=d(x_{i},y_{j})$ for all $1\leq i\leq M, 1\leq j\leq N$ for simplicity. 
Consequently, the RD function in \eqref{rate_dis_tor_def} can be 
equivalently written as:
\begin{subequations} \label{rd_model}
\begin{align}
&\min_{w_{i j}\ge0} & \sum_{i=1}^{M} \sum_{j=1}^{N} (w_{i j}p_{i}) \left[\log w_{i j}-\log \left(\sum_{i=1}^{M} w_{i j} p_{i}\right)\right] \label{rd_model_a} \\
&\text{ s.t.} &\sum_{j=1}^{N} w_{i j}=1, ~ \forall i=1,\cdots,M,\label{rd_model_b} \\
& &\sum_{i=1}^{M}\sum_{j=1}^{N} w_{i j} p_{i} d_{ij} \leq D. \label{rd_model_c}
\end{align}
\end{subequations}

Here, we notice the similarity between \eqref{rd_model} and the regularized OT problems. 
Specifically, the objective function in \eqref{rd_model_a} partially corresponds to the entropy regularization, \eqref{rd_model_b} is related to the marginal distribution. 
Hence, we are inspired to solve the problem from the OT approach.
However, the above problem is not a standard OT form, since the formulation \eqref{rd_model} lacks restrictions on the marginal distribution of the other side.
Moreover, the above optimization problem is hard to solve directly, due to the log-sum term $\log (\sum_{i=1}^{M} w_{i j} p_{i})$ in \eqref{rd_model_a}.
In order to overcome these difficulties, we introduce the slackness variable $\{r_{j}\}$ and propose the CommOT model as follows: 
\footnote{{ 
Introducing the variable $\bdr$ in \eqref{rd_model} is a straightforward method one can easily perceive.
%
%
Indeed, some previous studies \cite{EM_Algorithm} have attempted forms similar to the CommOT model \eqref{CommOT_model}.
However, to the best of our knowledge, we have not seen any discussion from the perspective of the OT theory.
%
%
%
%
%
%
%
%
%
}}
\begin{subequations} \label{CommOT_model}
\begin{align}
&\min_{w_{i j}\ge0,\ r_{j}\ge0} & \sum_{i=1}^{M} \sum_{j=1}^{N} (w_{i j}p_{i}) \left[\log w_{i j}-\log r_{j}\right] \label{CommOT_model_a} \\
&\text{ s.t.} &\sum_{j=1}^{N} w_{i j}=1, ~ \sum_{i=1}^{M} w_{i j} p_{i}=r_{j}, ~ \forall i,j, \label{CommOT_model_b} \\
& &\sum_{i=1}^{M}\sum_{j=1}^{N} w_{i j} p_{i} d_{ij} \leq D, \quad \sum_{j=1}^{N} r_{j}=1. \label{CommOT_model_c}
\end{align}
\end{subequations}

Note that the CommOT model \eqref{CommOT_model} 
%
%
is in fact a regularized OT model with an additional distortion threshold in communication theory, when we fix $\bdr$ as one-side marginal distribution.
In this way, we have the motivation to use the Sinkhorn algorithm in OT problems for high efficiency.
However, contrary to classical OT problems, the CommOT model has only one side of marginal distribution fixed, while the OT problem has two sides of fixed marginal distribution.
Therefore, the Sinkhorn algorithm can not be applied directly. 
To deal with this problem, we update the variables $\bdw$ and $\bdr$ alternatively to maintain the OT structure in the CommOT model.
In the next section, we modify the Sinkhorn algorithm, so that it is suitable for the computation of our proposed CommOT model.
%


\section{Alternating Sinkhorn Algorithm}  \label{sec_sinkhorn_3}
In order to solve the CommOT model \eqref{CommOT_model}, we first introduce dual variables $\bda\in\mbbR^{M}, \bdb\in\mbbR^{N}$, $\lambda\in\mbbR^{+}$ and $\eta\in\mbbR$, and then the Lagrangian function of problem \eqref{CommOT_model} is given by: 
\vspace{-.05in}
\begin{equation}\label{Lagrangian}
\begin{aligned}
\mathcal{L}&(\bdw, \bdr; \bda, \bdb, {\lambda}, {\eta})=\sum_{i=1}^{M}\sum_{j=1}^{N} w_{i j}p_{i} \left[ \log w_{ij}-\log r_{j}\right] \\
&+\sum_{i=1}^{M} \alpha_{i} \left(\sum_{j=1}^{N} w_{ij}-1\right)+\sum_{j=1}^{N}\beta_{j}\left(\sum_{i=1}^{M}w_{ij}p_{i}-r_{j}\right) \\ 
&+{\lambda}\left(\sum_{i=1}^{M}\sum_{j=1}^{N} w_{i j} p_{i} d_{ij}-D\right)+ {\eta}\left(\sum_{j=1}^{N} r_{j}-1\right).
\end{aligned}
\end{equation}
On this basis, our key ingredient is to take the derivatives of $\bdw$ and $\bdr$ to get the dual problem, and then update the dual variables alternatively, as described below:
%
%
\begin{itemize}
    \item[A.] Fix $\bdr$ as constant parameters, then update $\bdw$ and associated dual variables $\bda,\bdb,\lambda$; see Subsection \ref{subsec:w};
    \item[B.] Fix $\bdw$ as constant parameters, then update $\bdr$ and associated dual variable $\eta$; see Subsection \ref{subsec:r}.
\end{itemize}
%
%
%
A notable advantage of this procedure is that when $\bdr$ is fixed as constant parameters, the dual variables $\bda,\bdb$ can be updated by the Sinkhorn algorithm and $\lambda$ can also be updated easily; 
when $\bdw$ is fixed as constant parameters, the dual variable $\eta$ can also be updated efficiently by one-dimensional monotonic line search with only a few iterations.

\subsection{Updating $\bdw$ and its Dual Variable} \label{subsec:w}
Taking the derivative of $\mathcal{L}\left(\bdw, \bdr; \bda, \bdb, \lambda, \eta\right)$ with respect to the primal variable $\bdw$, one obtains
\begin{equation*} \label{dW}
\frac{\partial \mathcal{L}}{\partial w_{ij}}=p_{i}\left(1+\log w_{i j}-\log r_{j}\right)+\alpha_{i}+p_{i} \beta_{j}+\lambda p_{i} d_{ij},
\end{equation*}
and it further yields the representation of $\bdw$ by dual variables
\begin{equation} \label{update_W}
w_{ij}=\exp(-{\alpha_{i}}/{p_{i}}-\beta_{j}-\lambda d_{i j}-1+\log r_{j}).
\end{equation}
Denote $\phi_{i}=\exp\left(-{\alpha_{i}}/{p_{i}}-{1}/{2}\right)$, $\psi_{j}=\exp\left(-\beta_{j}-{1}/{2}\right)$ and $K_{ij}=\exp \left(-\lambda d_{i j}\right)$ for simplicity, then \eqref{update_W} yields
\begin{equation*}
w_{ij}=\phi_{i}\exp \left(-\lambda d_{i j}\right)\psi_{j}r_{j}=\phi_{i}K_{ij}\psi_{j}r_{j}.
\end{equation*}
Substituting the above formula into \eqref{CommOT_model_b}, we get
\begin{equation} \label{ot_cont}
\phi_{i} \sum_{j=1}^{N} K_{ij} \psi_{j} r_{j}=1, \quad
\psi_{j} r_{j} \sum_{i=1}^{M} K_{ij} \phi_{i}p_{i}=r_{j}. 
\end{equation}
By eliminating $r_{j}$ on both side of the \eqref{ot_cont}, we can alternatively update $\psi_{j}$ and $\phi_{i}$ by the Sinkhorn algorithm as follows: 
\begin{equation}\label{ot_sinkhorn} 
\psi_{j} \leftarrow 1 \bigg/ \sum_{i=1}^{M} K_{ij} \phi_{i} p_{i}, \quad \phi_{i} \leftarrow 1 \bigg/ \sum_{j=1}^{N} K_{ij} \psi_{j} r_{j}.
\end{equation}
Next, taking the derivative of $\mathcal{L}\left(\Tilde{\bdw}, \bdr^{n}; \bda, \bdb, \lambda, \eta^{n}\right)$ with respect to $\lambda$, we have the following condition for $\lambda\in\mbbR^{+}$
\begin{equation}\label{G_def}
G(\lambda) \triangleq \sum_{i=1}^{M}\sum_{j=1}^{N} d_{ij} p_{i}\phi_{i} \mathrm{e}^{-\lambda d_{i j}} \psi_{j}r_{j}-D=0.
\end{equation}
Note that the derivative of $G(\lambda)$ is negative, 
\begin{equation*}
G^{\prime}(\lambda)=-\sum_{i=1}^{M} \sum_{j=1}^{N} \phi_{i} \psi_{j} p_{i} r_{j} d_{i j}^{2} \mathrm{e}^{-\lambda d_{i j}}<0,
\end{equation*}
and hence $G(\lambda)$ is monotonic.
Therefore, we are to update the dual variable $\lambda\ge 0$ by finding the root of $G(\lambda)$ with Newton's method {with a few iterations}.
%
%
{Moreover, the feasibility of the Sinkhorn iteration can be guaranteed according to a similar discussion as in the last paragraph of Section III in \cite{ye2022optimal}.}

\subsection{Updating $\bdr$ and its Dual Variable} \label{subsec:r}

Taking the derivative of $\mathcal{L}\left(\bdw, \bdr; \bda, \bdb, \lambda, \eta\right)$ with respect to the primal variable $\bdr$ leads to the following equation: 
\begin{equation*}
\frac{\partial \mathcal{L}}{\partial r_{j}}=-\sum_{i=1}^{M} w_{i j}p_{i} \frac{1}{r_{j}}-\beta_{j}+\eta,
\end{equation*}
which yields a representation of $\bdr$ by dual variables $\bdb$ and $\eta$, 
\begin{equation}\label{from_of_r}
r_{j} = \left(\sum_{i=1}^{M} w_{i j}p_{i} \right)\bigg/\left(\eta-\beta_{j}\right).
\end{equation}
Substituting \eqref{from_of_r} into the equality constraint in \eqref{CommOT_model_c}, we have
%
%
\begin{equation}
F(\eta) \triangleq \sum_{j=1}^{N}\left[\left(\sum_{i=1}^{M} w_{i j}p_{i} \right)\bigg/\left(\eta-\beta_{j}\right)\right]-1=0. \label{F_def}
\end{equation}
Note that the derivative of $F(\eta)$ is negative, 
\begin{equation*} 
F^{\prime}(\eta) = -\sum_{j=1}^{N}\left[\left(\sum_{i=1}^{M} w_{i j}p_{i} \right)\bigg/\left(\eta-\beta_{j}\right)^{2}\right]<0,
\end{equation*}
and hence $F(\eta)$ is monotonic in each monotone interval.
In fact, we have $\eta > \beta_j,\forall j$ according to \eqref{from_of_r} and $r_{j}\ge 0$. 
Considering $\sum_{i}w_{i j}p_{j} > 0$, it is clear that $F(\eta) > 0$ for $\eta = \max_{j}(\beta_{j}) + \varepsilon$. Thus the rational function $F(\eta)$ has exactly one root in $(\max_{j}(\beta_{j}), +\infty)$. 
Therefore, we can simply adopt the Newton’s method {with a few iterations} to find the root of $F(\eta)$ in the interval $(\max_{j}(\beta_{j}), +\infty)$.
%

%
%

\vspace{+.1in}
To summarize, we update the dual variables $\bdb, \bda, \lambda$, and $\eta$ in an alternating manner and also update the variables $\bdw$ and $\bdr$ during the process.
For clarity, the pseudo-code is presented as follows.
\begin{algorithm}[H]
	\caption{Alternating Sinkhorn (AS) Algorithm}
	\label{alg:OT_rdf}
	\begin{algorithmic}[1]
		\REQUIRE Distortion measure $d_{ij}$, marginal distribution $p_{i}$, \\ maximum iteration number $max\_iter$.
		\ENSURE Minimal value $\sum_{i=1}^{M} \sum_{j=1}^{N} (w_{i j}p_{i}) \left[\log w_{i j}-\log r_{j}\right]$.
		\STATE \textbf{Initialization:} $\bm{\phi} = \mathbf{1}_{M}, \bm{\psi} = \mathbf{1}_{N}, \lambda=1, r_{j}=1/N$ ;
		\STATE Set $K_{ij} \gets \exp(-\lambda d_{ij})$
		\FOR{$\ell = 1 : max\_iter$}
		\STATE $\psi_{j} \gets 1/\sum_{i=1}^{M}K_{ij}\phi_{i}p_{i}, \quad j=1,\cdots,N$
		\STATE $\phi_{i} \gets 1/\sum_{j=1}^{N}K_{ij}\psi_{j}r_{j}, \quad i=1,\cdots,M$
		\STATE Solve $G(\lambda) = 0$ for $\lambda\in\mbbR^{+}$ with Newton's method
		\STATE Update $K_{ij} \gets \exp(-\lambda d_{ij})$ and $w_{ij}\gets\phi_{i}K_{ij}\psi_{j}r_{j}$
		\STATE Solve $F(\eta) = 0$ for $\eta\in\mbbR$ with Newton's method
		\STATE Update $r_{j}\gets\left(\sum_{i=1}^{M} w_{i j}p_{i} \right)\Big/\left(\eta-\beta_{j}\right)$ 
		\ENDFOR
		\STATE \textbf{end}
		\RETURN $\sum_{i=1}^{M}\sum_{j=1}^{N} \left(\phi_{i}p_{i}K_{ij}\psi_{j}r_{j}\right)\left[\log \left(\phi_{i}K_{ij}\psi_{j}\right)\right]$
	\end{algorithmic}
\end{algorithm}

\subsection{Relation with BA Algorithm}\label{subsec:differ_with_BA}
{
%
%
The AS algorithm solves the RD function with the CommOT model \eqref{CommOT_model}, whereas the BA algorithm uses model \eqref{rd_model}.
These two models are equivalent but in different forms.
The AS algorithm uses $r_{j}=\sum_{i=1}^{M} w_{i j} p_{i}$ as a marginal distribution constraint in the CommOT model \eqref{CommOT_model} to obtain an OT structure, 
while the BA algorithm only adopts this constraint when minimizing $r$ instead of directly introducing it in model \eqref{rd_model}.
%
%
}

{
Similar to the proof in \cite{book_BA} (see its Section 9.2.2 for details), it is easy to show that $\lambda$ in the AS algorithm is the slope of the RD curve, which corresponds to the annealing parameter in the BA algorithm.
Specifically, the AS algorithm updates the dual variable $\lambda$ in every iteration by Newton's method with only a few steps since we observe the nice monotonic property of $G(\lambda)$ in \eqref{G_def}. 
On the other hand, the BA algorithm fixes its parameter $\lambda$ as a constant throughout the iterations. Hence it cannot compute the rate $R$ directly for a given distortion $D$.
%
%
%
%
%
%
}

\section{Numerical Results and Discussions} \label{sec_num_4}
This section evaluates the efficiency of the proposed CommOT framework using the Alternating Sinkhorn (AS) algorithm and compares results with the BA algorithm.

\vspace{+.1in}
All the experiments are conducted on a PC with 8G RAM, and one Intel(R) Core(TM) Gold i5-8265U CPU @1.60GHz.
\subsection{CommOT Model over Classical Distributions}
This subsection computes the RD function of three classical distributions, \thatis, 
the binary source with Hamming distortion, the Gaussian source with squared-error distortion, and the Laplacian source with $L_1$ norm distortion.
The explicit expressions of the RD function in these cases can be found in \cite{book_element,berger1971} (see Section 4.3 and Section 10.3 respectively).
For Gaussian source and Laplaician source, we truncate the sources into an interval $[-M,M]$ and then discretize the interval by a set of uniform grid points $\{x_{i}\}_{i=1}^{2N+1}$, \thatis,
\begin{equation*}
x_{i}=-M+(i-1) \cdot \delta,~\delta={M}/{N},~i=1,\cdots,2N+1.
\end{equation*}
We further denote the discrete distribution $p$ of the sources by
\begin{equation}
p_{i} = F(x_{i+1})-F(x_{i}),\quad i=1,\cdots,2N,
\end{equation}
where $F(x)$ represents the distribution of the Gaussian source or the Laplacian source.
For the binary source, we can directly compute the result, since its distribution is discrete. 
\vspace{-.1in}

\begin{figure}[H]
    \centerline{\includegraphics[width=0.48\textwidth]{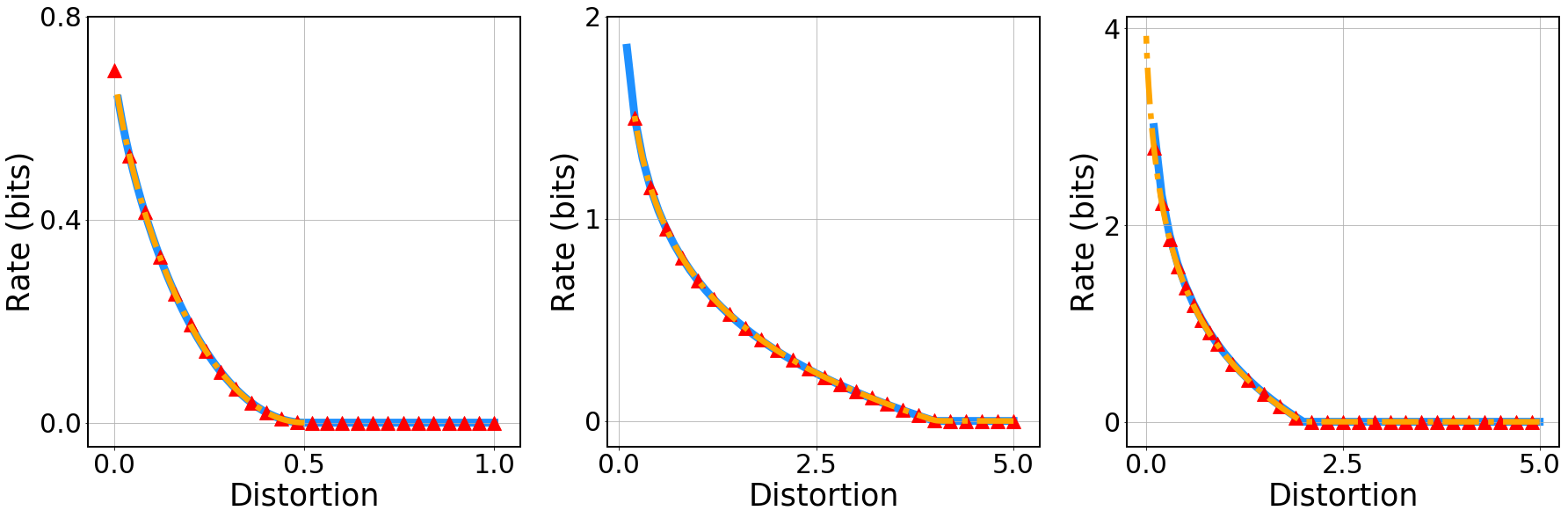}}
    \caption{The RD function obtained by the AS algorithm (red triangles), the BA algorithm (orange dashed lines), and the analytic expression (blue solid lines) under different sources, including the binary source of $p = 0.5$ (Left), the Gaussian source of $M=8,\delta=0.5,\sigma=2$ (Middle), and the Laplacian source of $M = 14, \delta=0.2, \sigma=2$ (Right), where $\sigma$ is the scale parameter in the Gaussian and Laplacian distribution. 
    %
    %
    For the curve plotted by the AS algorithm, the distortion $D$ is chosen as an arithmetic sequence of $25$ elements and we compute the result according to every distortion $D$. For the curve plotted by the BA algorithm, the slope $\lambda$ is chosen as an arithmetic sequence of $1000$ elements and we compute the result according to every slope $\lambda$.}
    \label{Fig:binary}
\end{figure}

\vspace{-.1in}
In Fig. \ref{Fig:binary}, we plot the curve of the RD functions given by the AS algorithm, the BA algorithm, and the analytic expression.
%
%
As it is shown in the above figure, our AS algorithm perfectly matches the analytic expression of the RD function as well as the BA algorithm in all three cases.
%
%
Thus, the accuracy of the AS algorithm is illustrated. 
Also, it is worth mentioning that even though the choice of $\lambda$ is very dense, the RD curve plotted by the BA algorithm is still sparse for large distortion regimes, especially in the interval $[1,4]$ of the Gaussian case.
This phenomenon shows that it is difficult for the BA algorithm to obtain the RD function with a specific distortion threshold.
%

\vspace{-.1in}
\begin{figure}[H]
    \centerline{\includegraphics[width=0.45\textwidth]{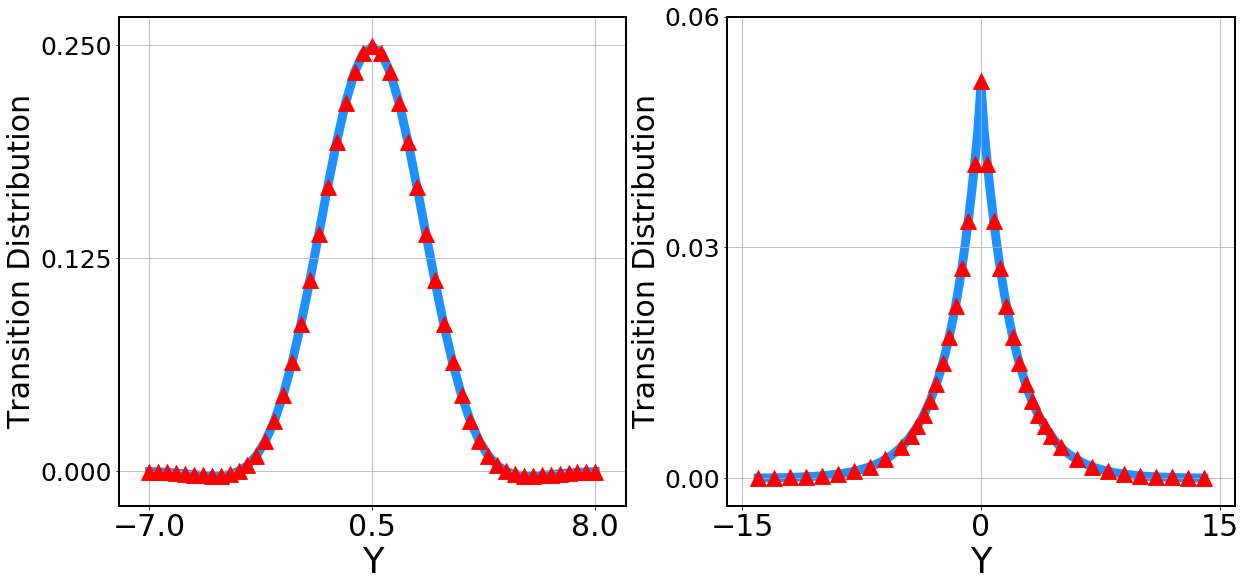}}
    \caption{The marginal distribution of Y rebuilt by the AS algorithm (red triangles) and the analytic expression (blue solid lines), under the Gaussian source of $M=8,\delta=0.5,\sigma=2,D=1$ (Left), and the Laplacian source of $M = 14,\delta= 0.2,\sigma=2,D=0.1$ (Right).}
    \label{Fig:Distribution_on_Y}
\end{figure}

\vspace{-.1in}
Besides, for the aforementioned Gaussian and Laplacian cases, we consider the rebuilt marginal distribution of $Y$ with a given distortion $D$, in Fig. \ref{Fig:Distribution_on_Y}.
The marginal distribution given by the AS algorithm perfectly matches the analytic expression in the above two cases \cite{berger1971}, while the BA algorithm can hardly handle this task \cite{book_BA} due to the aforementioned reason.
To further illustrate the accuracy and efficiency of the AS algorithm and the CommOT model, we compare the result with the BA algorithm as baseline. 
The averaged computing time and the averaged difference of the optimal values between the two algorithms are listed in Table I.
To reduce the influence of noise, we repeat each experiment for $100$ times. 
Here, we only change the distortion $D$ to compare the results of the two algorithms.
From the table, we can see the optimal values obtained by the two algorithms are almost the same (with a difference less than $1e$-$7$).
But our proposed AS algorithm has a visible advantage in computing time.

\vspace{-.15in}
\begin{table}[ht] \label{table_compare} 
	\renewcommand\arraystretch{1.4}
	\centering
	\caption{Comparison between the AS algorithm and the BA algorithm. } 
	\setlength{\tabcolsep}{1.2 mm}{
		\begin{tabular}{c|c|c|c|c|c}
			\toprule
			\multirow{2}{*}{} & \multirow{2}{*}{($D$, Slope $\lambda_{D}$)} & \multicolumn{2}{c|}{Time (s)} & {Ratio} & {Error} \\
			\cline{3-4} \cline{5-6}
			 &  & $t_{AS}$ & $t_{BA}$ & Speed-up & $|$AS$-$BA$|$\\ 
			\hline 
			\multirow{2}{*}{Binary}
			& $(0.1,2.1972)$ & $0.0021$ & $0.0388$ & $18.15$ & $1.05\!\times\!10^{-15}$ \\
			& $(0.4,0.4055)$ & $0.0022$ & $0.0330$ & $15.30$ & $4.12\!\times\!10^{-15}$ \\
			\hline
			\multirow{2}{*}{Gaussian} 
			& $(0.5,1.0000)$ & $0.2429$ & $3.0951$ & $12.74$ & $1.32\!\times\!10^{-8}$ \\
			& $(1.0,0.5000)$ & $ 0.2768$ & $ 3.0804$ & $11.13$ & $2.44\!\times\!10^{-8}$ \\
			\hline
			\multirow{2}{*}{Laplacian} 
			& $(0.5,1.9499)$ & $1.4710$ & $21.4264$ & $14.57$ & $1.32\!\times\!10^{-8}$ \\
			& $(1.0,0.9931)$ & $1.6287$ & $30.0881$ & $18.47$ & $2.01\!\times\!10^{-8}$ \\
			\bottomrule
	\end{tabular}}
 
\vspace{+.03in}

\footnotesize{{Notes: a) Columns 3-4 are the averaged computing time of the two algorithms, and columns 5-6 are the speed-up ratio and average difference between the AS and BA algorithm. b) The BA algorithm cannot compute the rate directly with a given distortion $D$, and hence we adaptively search the corresponding slope $\lambda_D$ to ensure accuracy. It generally takes about $30\sim 80$ to search for a suitable slope $\lambda_D$. c) The maximum number of iterations for both algorithms is 1000 in all tested cases.}}

\end{table}

\vspace{-.15in}

\subsection{Convergence Behaviour and Algorithm Verification}
In this subsection, we study the convergence of the AS algorithm by considering the residual errors of the Karush-Kuhn-Tucker (KKT) condition of \eqref{CommOT_model}.
Here, we define the absolute residual errors $r_{\psi},r_{\phi},r_{\lambda},r_{\eta}$ as:
\begin{equation*}
\begin{aligned}
&r_{\psi}\!=\!\sum_{j=1}^{N}\left|\psi_{j}\sum_{i=1}^{M}K_{ij}\phi_{i}p_{i}-1\right|,~ r_{\phi}\!=\!\sum_{i=1}^{M}\left|\phi_{i}\sum_{j=1}^{N}K_{ij}\psi_{j}r_{j}-1\right|, \\
&r_{\lambda} \!=\! \left|\lambda (\sum_{i=1}^{M} \sum_{j=1}^{N} \phi_{i} \psi_{j} p_{i} r_{j} d_{ij} K_{i j}-D)\right|, ~~ r_{\eta} \!=\! \left|\sum_{j=1}^{N} r_{j}-1\right|.
\end{aligned}
\end{equation*}

\vspace{-.2in}
\begin{figure}[H]
 \centerline{\includegraphics[width=0.5\textwidth]{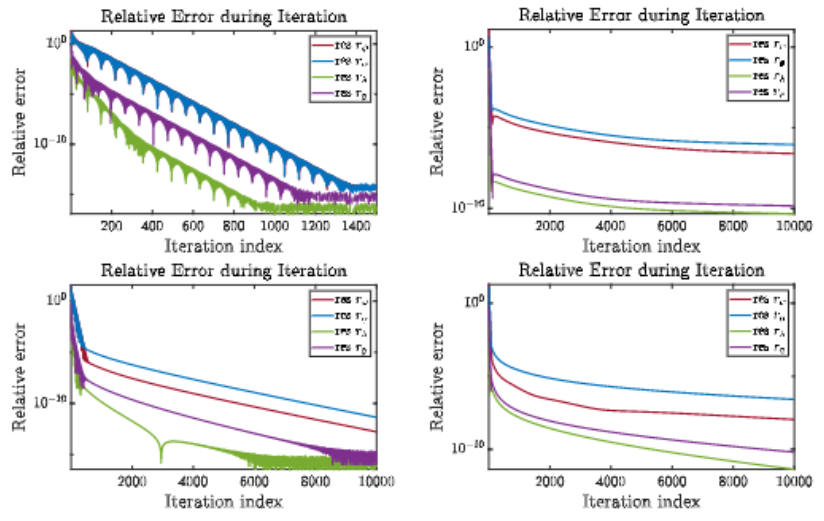}}
 \caption{The convergent trajectories of the residual error for $r_{\psi}$ (Red), $r_{\phi}$ (Blue), $r_{\lambda}$ (Green) and $r_{\eta}$ (Purple) for the Gaussian source, where distortion $D =  0.1$ (Upper Left), $0.5$ (Upper Right), and the Laplacian source, where distortion $D =  0.5$ (Lower Left), $1.0$ (Lower Right).} 
 \label{res}
 \end{figure}
%
%
%
%
%
%
%
%

%
In Fig. \ref{res}, we plot the convergent trajectories of the residual errors of the Gaussian source and the Laplacian source with respect to iteration steps.
In these two experiments, we set the same parameters as in Fig. \ref{Fig:binary} and the maximum number of iterations is $10000$.
As it can be seen in Fig. \ref{res}, for different distortions, the AS algorithm will converge to $1e$-$7$.
Also, we note that the convergence of the AS algorithm can be influenced by the choice of distortion $D$, {which is similar to the discussion of convergence near bifurcation points in \cite{agmon2022root}.}

\subsection{Bifurcation Problems and Linear Segment Support}

\begin{figure}[H]
	\centerline{\includegraphics[width=0.41\textwidth]{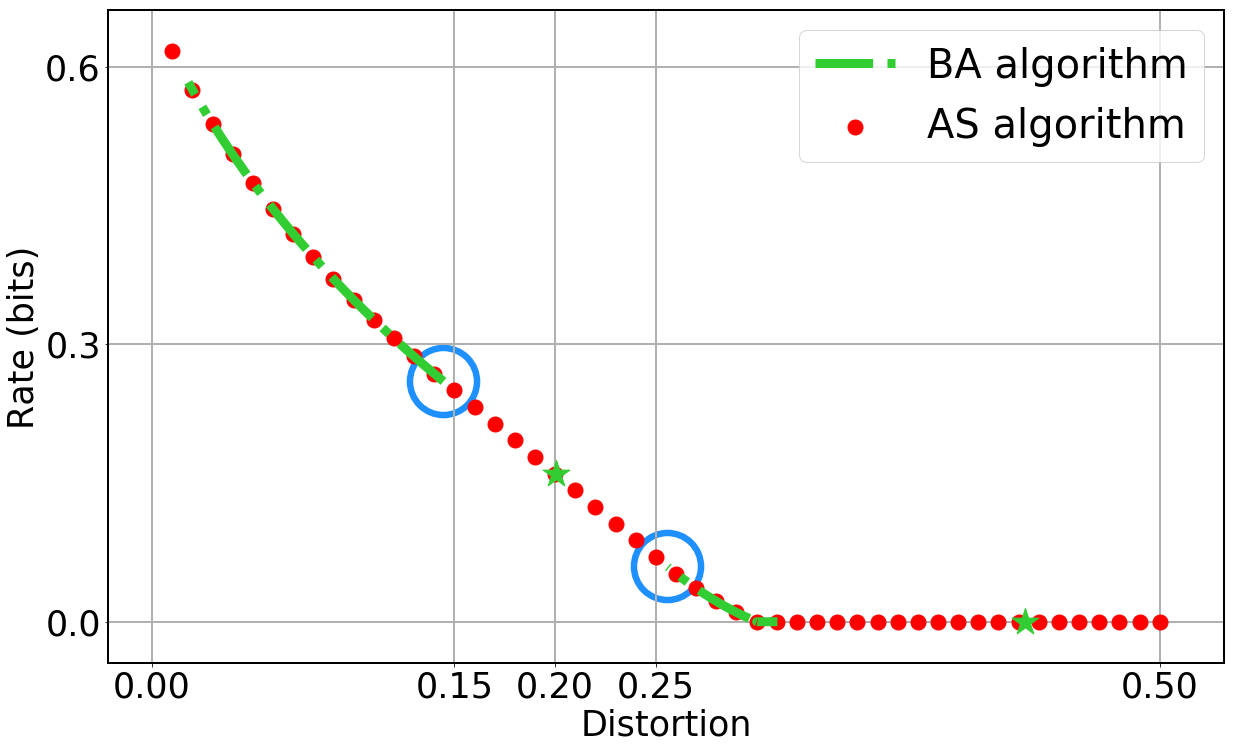}}
	\caption{The RD function obtained by the AS algorithm (red points) and the BA algorithm (green dashed lines and stars) under the support-switching and a cluster-vanishing bifurcation case.}
	\label{Fig:baff}
\end{figure}

In this subsection, {we consider the case with bifurcation structure and linear segment support \cite{agmon2021critical,agmon2022root}. 
Here we define the distortion measure and the marginal distribution as in \cite{berger1971},}
\begin{equation*}
d(x, \hat{x})=\left(\begin{array}{ccc}
1 & 0 & 0.3 \\
0 & 1 & 0.3
\end{array}\right) \text { and } p_{X}=(0.4,0.6).
\end{equation*}
This problem exhibits two bifurcations, at $D_{1}$ and $D_{2}$ (marked with blue cycle in Fig. \ref{Fig:baff}). 
Here distortion $D_{1}$ is around $0.14$ and $D_{2}$ is around $0.25$.
The curve with the linear segment is between the two distortion points $D_{1}$ and $D_{2}$.
The linear segment can be explained by a support-switching bifurcation between two sub-optimal RD curves (see \cite{agmon2022root} Section 6.5).
As shown in Fig. \ref{Fig:baff}, the AS algorithm can output the accurate results. 
In particular, in the linear segment part, there is only one unique slope value, and thus the BA algorithm cannot output the whole line with high accuracy \cite{agmon2022root}.
However, in the AS algorithm, we update the slope $\lambda$ with distortion $D$ fixed, so that this issue can be handled easily.

\section{Conclusion} \label{sec_conclu_5}
This paper presents a new framework for computing the rate distortion function. 
We give a general optimal transport framework named CommOT to formulate this problem.
%
To solve the CommOT model, we propose the AS algorithm and test it under some classical sources, e.g. binary, Gaussian, and Laplacian sources.
The numerical results based on the AS algorithm agree well with the analytical solution in all those cases. 
Based on our proposed model and algorithm, we expect that the CommOT framework can be applied to various theoretical bounds in communication theory, such as the rate distortion perception trade-off \cite{blau2019perception} and the information bottleneck problem \cite{6795936}.
%

\bibliographystyle{bibliography/IEEEtran}
\bibliography{bibliography/RD_REF}

\end{document}